# Study of thickness dependent density in ultrathin water soluble polymer films

Mojammel H. Mondal, M. Mukherjee,

*Surface Physics Division, Saha Institute of Nuclear Physics, 1/AF, Bidhannagar, Kolkata-64, India*

K. Kawashima, K. Nishida, T. Kanaya

*Institute for Chemical Research, Kyoto University, Uji, Kyoto-fu, 611-0011, Japan*

ABSTRACT: Density of the polyacrylamide ultrathin films has been studied using X-ray reflectivity technique. Two sources (one powder and another aqueous solution) of polyacrylamide were used to prepare spin coated films on silicon substrate. Light scattering measurements show that the polymer chains were unentangled in a concentrated (4 mg/ml) as well as in a dilute (2 mg/ml) solution prepared from the powder, whereas the solution (4 mg/ml) prepared by diluting the solution source shows entangled chain morphology. Three sets of films of different thicknesses were prepared using the three solutions by spin coating on silicon substrates. Comparison of X-ray reflectivity data for as prepared and dry films reveals that the shrinkage of the films decreases with increasing thickness. Average electron densities of the films were found to follow a trend of higher density for thinner films with a maximum increase of about 12% compared to the bulk. The densities of all the films irrespective of the nature of entanglement and concentration of their source were found to increase with spin speed of coating and attain saturation at higher speed. Absence of correlation between shrinkage and density data and the fact that the densities of all the films follow a master curve irrespective of their origin suggest that the higher density of the films result from the higher orientation of chains as a consequence of an interplay between stretching and stronger attractive interactions of polar nature.

**Introduction**

Ultrathin polymer films have become the subject of extensive research to understand fundamental physical properties of these materials in confined geometry that are important for applications in nanotechnology. Either supported or freely standing films have been investigated extensively using different experimental techniques including ellipsometry, [1] Brillouin light scattering, [2] positron annihilation lifetime spectroscopy [3,4] or X-ray and neutron reflectivity [5-9] to study the properties like glass transition temperature, thermal contraction or expansion behavior, wetting, solvent absorption and swelling to name a few. One important fundamental property of polymer thin films is the mass density on which many of the other properties like mechanical or glass transition temperature $T_g$ for example, depend.

Reduction of the glass transition temperature with film thickness has been observed by several authors. [2, 9-11] The reason behind such behavior can be found in a simulation study [12] that shows the enhancement of the mobility of the polymer chains due to the lowering of density of the polymer films at polymer vacuum interface. However, until now there is no direct experimental proof of the prediction and it is still a matter of debate whether or not the density of a polymer in an ultrathin film is different from its density in bulk. Many authors have studied the density of thin films and have arrived in conflicting conclusions. Reiter [13] has used X-ray reflectivity to infer that polystyrene films have lower mass density than bulk for the films thinner than the radius of gyration $R_g$. Neutron reflectivity measurements of very thin duterated poly(methyl methacrylate) films show increased [14] as well as decreased [15] mass density. Recently neutron reflectivity has been used to demonstrate nearly unaltered [11] mass densities for the thin polystyrene films compared to their bulk values. Variations of densities only at the film-air or film-substrate interfaces have been predicted by several authors. X-ray reflectivity has been used to predict lower density for poly(methyl methacrylate) and polystyrene [16] and higher density for poly(4-vinylpyridine) related to the existence of weak or strong adsorption of the polymers with the substrate. Positron annihilation spectroscopy has been employed to show the presence of density gradient at the film-air interface [17] or to have broader free-volume and hole distribution near the top surface than in the bulk [18] for polystyrene films.

It is clear from the above discussion that all the studies on the density of thin polymer films are performed with polymers that are soluble in organic solvents, as these polymers play important roles in various applications such as coating of television screens, optical devices, and magnetic storage disks. [19, 20] Water soluble polymers on the other hand constitute an important class of materials for their high viscosity, owing to the hydrogen bonding capabilities with easy to use, cheap and health hazard free solvent. Recently there has been an escalated interest in the fields of water soluble polymer thin films. Authors have discussed structural aspects of pure polyacrylamide and nanocomposites ultrathin films. [21-24] Swelling dynamics of these films in presence of saturated vapor also have been discussed by these authors. However, there was no report in the literature on the density of the thin water soluble polymer films until recently [23] where the authors have reported relatively large change in the mass density of the polyacrylamide films as a function of thickness in contrary to the prediction of a small or nearly no change of density for films of polymers soluble in organic solvents.

Motivated by the contrasting predictions of density of thin films compared to the bulk values between the polymers soluble in organic solvent and the water soluble one we have performed X-ray reflectivity study of spin coated polyacrylamide ultrathin films prepared from aqueous solutions of various morphological conditions, concentrations and spin speeds to compare their densities. Light scattering measurements were performed to study the morphology of the different solutions from which films have been prepared. Here we have discussed the possible mechanism for the density variation of the films with respect to starting solution morphology, film thickness and spin speed of coating.

**Experimental Details**

**Sample Preparation.** A powder and an aqueous solution (1%, 10 mg/ml) of high molecular weight polyacrylamide (Supplied by Polysciences, USA) were taken as starting materials for the experiment. Two solutions with concentration 2 mg/ml and 4 mg/ml were prepared from the powder source (S1, molecular weight 5-6 ×$10^6$). Another solution (4 mg/ml) was prepared from the solution source (S2, molecular weight 5 ×$10^6$). The solutions are denoted as solutions A, B and C respectively. Three sets of films from the three solutions were prepared on silicon substrate by spin coating method. The films prepared from the solutions A, B and C were designated as A, B, C films respectively. During the spinning, clean and warm (60°C) air was flown gently over the sol using a homemade arrangement to facilitate faster evaporation of water. Before coating, silicon wafers were cleaned by RCA cleaning method, where the wafers were boiled at 100°C for about 15 minutes in a solution of $H_2O$, $NH_4OH$ and $H_2O_2$ (volume ratio, 2:1:1). The wafers were then rinsed with Millipore water. Apart from cleaning, this treatment enhances the hydrophilicity of the silicon surface by introducing –OH dangling bonds on the surface which helps better attachment of the water soluble polymers. Films of different thicknesses were prepared applying different spinning speeds ranging from 500 to 4000 r.p.m.

**Light scattering.** To observe the correlation between the chain conformations in solution and the behavior of the thin films prepared by them, it was necessary to study the morphology of the starting solutions. Static (ALV-5000) and Dynamic (Zetasizer Nano-S, Malvern Instruments and ALV-5000) light scatterings were used to measure the scattering cross-section *I(q)* and the intensity-intensity time correlation (autocorrelation) functions $g_2(t)$ of the two source polymer solutions. The chain overlap concentration $C^*$ was found to lie in a range between $3M/(4\pi N_A R_g^3)$ to $M/(N_A R_g^3)$, where $M$ is molecular weight of the polymer, $N_A$ is the Avogadro number and $R_g$ is the radius of gyration of the polymer.[25] According to above conditions $C^*$ for our materials should have values in the range of 2-8 mg/ml. The light scattering measurements were performed with two different solutions having concentrations (0.5 mg/ml and 1 mg/ml) below $C^*$ for both the starting materials S1 and S2. The overall behaviors of each sample were similar for both the concentrations. In figure 1 we have shown the light scattering data for 0.5 mg/ml solutions of S1 and S2. Figure 1(a) shows the correlation functions obtained from the dynamic light scattering for both the samples. The data for S1 can be fitted well with a single exponential function featuring single decay behaviour in this sample whereas the data for S2 was fitted with a sum of three exponential decay functions showing multimodal decay with at least three distinct decay modes. The largest fraction of decay component with the intermediate relaxation time (~30 ms) for S2 was found to be two orders of magnitude larger than that (0.2 ms) of S1. The fact that the correlation function of the S2 can be described as a sum of three exponential functions suggests that this sample was dynamically heterogeneous compared to S1.

Figure 1(b) shows the static light scattering intensity for two samples (S1 and S2) as a function of momentum transfer $Q$. The data corresponding to S2 prepared from solution source shows very high intensity in the low $Q$ region suggesting heterogeneous structure in the sample which may be attributed to aggregation and/or gelation of polyacrylamide chains. The intensity for S1 prepared from powder source shows much lower intensity compared to S2 which signifies that the solution S1 prepared from powder source was rather homogeneous.

The static and dynamic light scattering results in general suggest that the solutions prepared from S2 (solution source) was more heterogeneous than that prepared from the powder source. It is expected that the chains of S2 was partially aggregated or cross-linked to give heterogeneous structure [25, 26] in the solution.

**Relative viscosity.** The ratios of the viscosities of water and the three solutions A, B and C used for spin coating were measured through efflux time. Fixed amount of solutions for all the samples were allowed to flow under the influence of gravity through a small orifice of an elongated tube and flow time was measured with a stop watch. The ratios of the viscosities obtained from the ratios of the efflux times for the three solutions were found to be $\eta_w : \eta_A : \eta_B : \eta_C$ = 1:1.16:1.58:1.7. The viscosity of solution C was higher than that of solution B despite having identical concentration in both. This behaviour supports the observation of partial cross linking in solution C observed from light scattering experiments described earlier.

**X-ray Reflectivity.** X-ray reflectivity is one of the best nondestructive methods to measure the structural aspects of thin polymeric films; here we have used this technique to study the thickness and electron density of polyacrylamide films prepared from the three solutions. X-ray reflectivity data were collected in our laboratory setup (Bruker AXS, D8 Discover) with CuKα radiation obtained from copper sealed tube anode (2.2 kW), followed by a Göbel mirror for focusing. Specular scans with identical incoming and outgoing angles for X-rays were taken as a function of momentum transfer vector $q$ normal to the surface ($q = (4\pi/\lambda) \sin\theta$, with $\theta$ the incident and the reflected angles of the X-ray and $\lambda$ = 1.54Å, the wavelength of the radiation). It was observed that spin coated polyacrylamide films retain internal strain in the as coated condition. All the films were swelled in saturated water vapor condition in a closed chamber at room temperature for twelve hours to release the strain before they were dried and stored in a desiccator. As the films of water soluble polymer are hygroscopic in nature, it was important to remove absorbed water molecules from the films in order to study their actual structure. At first, X-ray reflectivity data for all the films were collected at room temperature in a chamber continuously evacuated by a rotary pump. The corresponding thicknesses of the films were considered as initial film thickness. The films were subsequently heated at 105$^o$C for 45 min in vacuum and the reflectivity data at this temperature were collected in situ in vacuum. The total time at 105$^o$C including that of data acquisition was about three hours for all the films. The data were taken at high temperature as a precaution to avoid possibility of re-absorption of water vapor, through leakage, on reduction of temperature. From X-ray photoemission spectroscopy studies we found that drop cast thick polyacrylamide films heated above 100$^o$C under vacuum for one hour contains very small amount of water (one molecule of water for every 17 monomer units) and does not degas further in Ultra high vacuum (UHV) condition. The temperature applied for removal of the solvent was kept just above the boiling point of water in order to minimize structural modifications through the mobility of the chains as the objective of the study was to find the density of the films in the as cast condition.

To obtain information about the thickness and electron density of the films, the reflectivity data were analyzed using Parrat formalism [27] modified to include interfacial roughness. [28] For the analysis of the X-ray reflectivity data, the input electron density profiles were divided into several boxes of thickness equal or more than *2π/q$_{max}$* and the interfacial roughness were kept within 2-8 Å. During the analysis, the roughness of the polymer surface, the electron density, the thickness of the films and the roughness of the substrate were used as fitting parameters. Typical reflectivity data and the fitted profiles for a film treated at different temperatures have been shown in figure 2.

**Results and Discussion**

Reduction of film thickness was observed as the films were annealed at 105$^o$C. The change of thickness due to this annealing can be considered only due to the release of absorbed water from the free volume of the films as the temperature was far below the $T_g$ of the polymer. This quantity therefore measures the retention of solvent in the films. The presence of solvent is important for thin films as physical properties change due to its presence. [29-31] The questions relate to the quantity and the location of the retained solvent in the films. In figure 3 (a) and (b) we have plotted the absolute and relative change in thickness or shrinkage of the films given by $\Delta d_a = d_i - d_{105}$ and $\Delta d_r = \frac{(d_i - d_{105}) \times 100}{d_i}$ respectively for the three sets of films, as a function of their thicknesses at 105°C, where $d_i$ and $d_{105}$ denote the initial thickness and the thicknesses at 105°C respectively. It can be observed from figure 3a that the amount of retained water in the films increases with thickness of the films which indicates that water was stored in the entire film. In figure 3b, although few data points are somewhat scattered there is sufficient indication that relative change in thickness of the thinner films was more compared to the thicker ones. It may be noted that the shrinkage depends on the initial thickness $d_i$ of the films which has some dependence on the temperature and the humidity of the experimental room. As it was difficult to maintain the value of these parameters at constant level for several days during which the experiment was performed, deviation in some $d_i$ values are not unlikely. Many authors have shown that in case of thin films the solvent retains at the film substrate interface for the polymers that are soluble in organic solvents. [31-38] Contrary to these results, our earlier study on swelling of polyacrylamide films suggests that water resides in the entire film and concentration of the same increases towards the top layer [23] and the present shrinkage data are in agreement with our earlier observation.

In figure 4 we have plotted the thickness of the three sets of the films as functions of the speeds of spin coating, as obtained from the reflectivity measurements. It can be observed from the figure that the thickness versus spinning speed behavior for the three sets were similar with significant differences. For the films C the thickness of the films reduces monotonically with the spinning speed, whereas, for the films A and B the behavior does not appear to be very smooth, particularly at lower speeds. Although the concentrations of solution B and C are identical, the films C are much thicker compared to the films B. The smooth behaviour of thickness with spin speed and the higher thickness values for film C may be attributed to the higher viscosity of solution C compared to solution B. The spin coating process can be divided into three stages namely spin up, spin off and drying. Spin up is the first stage when the disc is accelerated to the final speed when the excess liquid is thrown off and a thin liquid film is created on the rotating disc very quickly in first one or two seconds. During the spin off stage the film is thinned due to a combination of convection and solvent evaporation. The centrifugal force drives the fluid radially towards the edge of the disk against the viscous resistance. This radial flow diminishes as the solvent evaporates and the viscosity is increased by several orders of magnitude. This stage of the coating process immensely influences the uniformity and the final thickness of the films [39] where viscosity of the liquid plays the central role. In case of solution C the higher viscosity acts as an impediment to the radial fluid flow resulting in higher thickness for films C compared to those from solution B for which the viscosity was considerably low. In the inset of figure 4 the logarithmic plot of the same data sets are presented. The slopes of the curves (*b*) that represent the power law behaviour [39] of the film thickness $h \propto \omega^{-b}$, with spin speed ω, of the three sets of data were found to be (A) 0.43 ± 0.09, (B) 0.57 ± 0.05 and (C) 0.52 ± 0.03. The numbers are in close agreement with the predicted value of 0.5 for Newtonian Fluids. [40-42] The only difference observed from the data of figure 4 between the films cast from uncrosslinked and the partial crosslinked solutions, the higher thickness of films C compared to those of films B, could be explained in terms of increased viscosity for solution C.

From X-ray reflectivity measurement one can obtain the electron density profile for a film along its depth, averaged over the probe size. In figure 5 we have plotted the average electron densities $\rho_{el}$ for all the films as a function of their thickness. The three sets of films are found to have similar average electron densities but with three different range of thicknesses. It can be seen from the figure that the electron densities of the polyacrylamide films increase with reduction of film thickness systematically for all the three sets of films. One may be inclined to explain the increase of film density in terms of the loss of solvent on annealing, particularly when one is biased by figure 3a where the absolute change in thickness or in other words solvent loss is found to increase with film thickness. However, in figure 3b one can find the relative change in thickness or the solvent loss per unit thickness, a more relevant parameter, shows a different behaviour. Moreover, in figure 3b careful observations show various types of changes such as decreasing, increasing or nearly constant trends within small ranges of successive data points. If the reason for change of density was the loss of solvent, it would be necessary that the change of density in figure 5 follow the variation of the data in figure 3b for different films. Absence of any such correlation between the data of figure 3b and figure 5 appears to discard the explanation of density change in terms of loss of solvent and necessitates an alternate explanation for the same.

In figure 6 we have plotted the average electron densities of the films against the spin speed of coating. It can be clearly observed from the figure that irrespective of the nature and concentration of the solutions used for preparing the films the electron densities increase with spinning speed and finally saturates to a value. An exponential fit to the data shown by the dashed line in the figure may be considered as a master curve to describe the behaviour of electron density of the films as a function of spinning speed of preparation. The maximum increase of electron density at saturation was about 12% with respect to the bulk electron density (0.42 Å$^{-3}$) of the polymer. It is interesting to note that the nature of density change with spin speed does not differentiate between partially entangled and unentangled solutions also. As a completeness of the study we have taken three annealed samples (film B) and further annealed them at the glass transition temperature of the polymer at 165$^{o}$C for 60 min under vacuum and X-ray reflectivity data of the films were taken in vacuum at 165$^{o}$C. The total time of annealing including the scan time was about three hours. The thickness and electron density of the films were found to remain nearly unchanged after annealing at $T_g$ as shown in table 1. This further indicates that the density of the dry films are determined by the initial condition at which the films are prepared and further chain movements at $T_g$ do not affect the morphology of the chains up to the extent of modifying their thickness or density. It appears from the data that the densities of the films are only dependent on the spin speed at which they are cast. It may be noted that all the films used in the present study were having thicknesses less than the radius of gyration ($R_g$) of the polymer (~ 100nm), thus presence of multiple layers of polymer coils in these films are unlikely. We therefore intend to believe that the phenomenon of increase in the electron density with reduction of film thickness or more fundamentally with increase of spin speed was the outcome of the higher in plane alignment of the polymer chains as a result of higher stretching. [23] It is known that during spin coating process various parameters come into play such as the solution viscosity, evaporation rate of the solvent, morphology of the substrate and the interaction between the substrate and the polymer. In the present case unlike the polymers that are soluble in organic solvents, the polymer-polymer, polymer-solvent and polymer-substrate interactions are strong due to the polar nature of the solvent and the polymer compared to the weak Van der Wall type interactions in the former. We believe that during spin coating of water soluble polyacrylamide films the chain morphology of the polymer coils are affected by the stretching as well as these strong attractive interactions and at higher spin speeds films are formed with enhanced in plane alignment of the chains resulting in reduction of free volumes and higher density

**Conclusions**

In conclusion, densities of polyacrylamide spin coated films prepared from entangled and unentangled solutions of various concentrations have been studied using X-ray reflectivity technique. A powder and an aqueous solution, as supplied, were used as starting materials. Two solutions (2 mg/ml (A) and 4 mg/ml (B)) prepared from the powder source and one (4 mg/ml (C)) from the aqueous solution source was used for the spin coating. Static and dynamic light scattering have been used to study the morphology of the polymer chains in the solutions used for spin coating. Light scattering results show unentangled chain conformation in solutions A and B whereas for solution C the chains have interchain entanglements. The relative viscosity of solution C was found to be larger than that of solution B although both have same concentration which also suggests that solution C has entanglement possibly due to partial cross linking within the chains. Three sets of spin coated films were prepared from the three solutions A, B and C at different speeds. X-ray reflectivity measurements of the as grown and dry films were performed under vacuum to study the structural aspects of the films. The thicknesses of the films were found to reduce after drying at $105^oC$ under vacuum. The relative shrinkage of the thinner films was larger compared to the thicker ones. The average electron densities of the films were found to have similar trend of higher density with thinner films with a maximum increase of about 12% compared to the bulk. The density of all the films was found to increase with spin speed of preparation to saturation irrespective of the nature and concentration of their source. Absence of correlation between the shrinkage and density data and the fact that that density of all the films follow a master curve irrespective of their origin suggest that the higher density of the films result from the higher orientation of chains due to stronger polymer-solvent, polymer-polymer and polymer-substrate interactions.

# Acknowledgements

Authors (M. H. Mondal and M. Mukherjee) thankfully acknowledge Prof. S. Hazra for extending the X-ray reflectivity facility for the study.
# References and Notes

(1) Keddie, J. L.; Jones, R. A. L.; Cory, R. A.; *Europhys. Lett.* **1994**, *27*, 59.

(2) Forrest, J. A.; Dalnoki-Veress, K.; Stevens, J. R.; Ducher, J. R.; *Phys. Rev. Lett.* **1996**, *77*, 2002.

(3) DeMaggio, G. B.; Frieze, W. E.; Gidley, D.; Zhu, W. M.; Hristov, H. A.; Fee, A. F.; *Appl. Phys Lett.* **1997,** *78*, 1524.

(4) Xie, L.; Demaggio, G.B.; Frieze W.E.; Devries, J.; Gidley, D.W.; Hristov, H.A.; Yee, A. F. *Phys. Rev. Lett.* **1995**, *74*, 4947.

(5) Wallace, W. E.; van Zanten, J. H.; Wu, W. L.; *Phys. Rev. E* **1995,** *52*, R3329.

(6) Orts, W. J.; van Zanten, J. H.; Wu, W.-L.; Satija, S. K.; *Phys. Rev. Lett.* **1993**, *71*, 867.

(7) Mukherjee, M.; Bhattacharya, M.; Sanyal, M. K.; Geue, T.; Grenzer, J.; Pietsch, U. *Phys. Rev. E* **2002**, *66*, 061801.

(8) Singh, A.; Mukherjee, M. *Macromolecules* **2003**, 36, 8728.

(9) Miyazaki, T.; Nishida, K.; Kanaya, T.; *Phys. Rev. E* **2004**, *69*, 061803.

(10) Forrest, J. A.; Dalnoki-Veress, K.; Stevens, J. R.; Dutcher, J. R. *Phys. Rev. E.* **1998**, *58*, 6109.

(11) Wallace,W. E. ; Beck Tan, N. C.; Satija, S. and Wu, W. L. *J. Chem. Phys.* **1998,** *108*, 3798.

(12) Mansfield, K.F. and Theodorou, D. N. *Macromolecules* **1991**, 24, 6283.

(13) Reiter, G. *Europhys. Lett* **1993**, *23*, 579.

(14) Fernandez, M. L.; Higgins, J. S.; Penfold, J. and Shackleton, C. S. Polym. Commun. **1990**, *26*, 252.
13

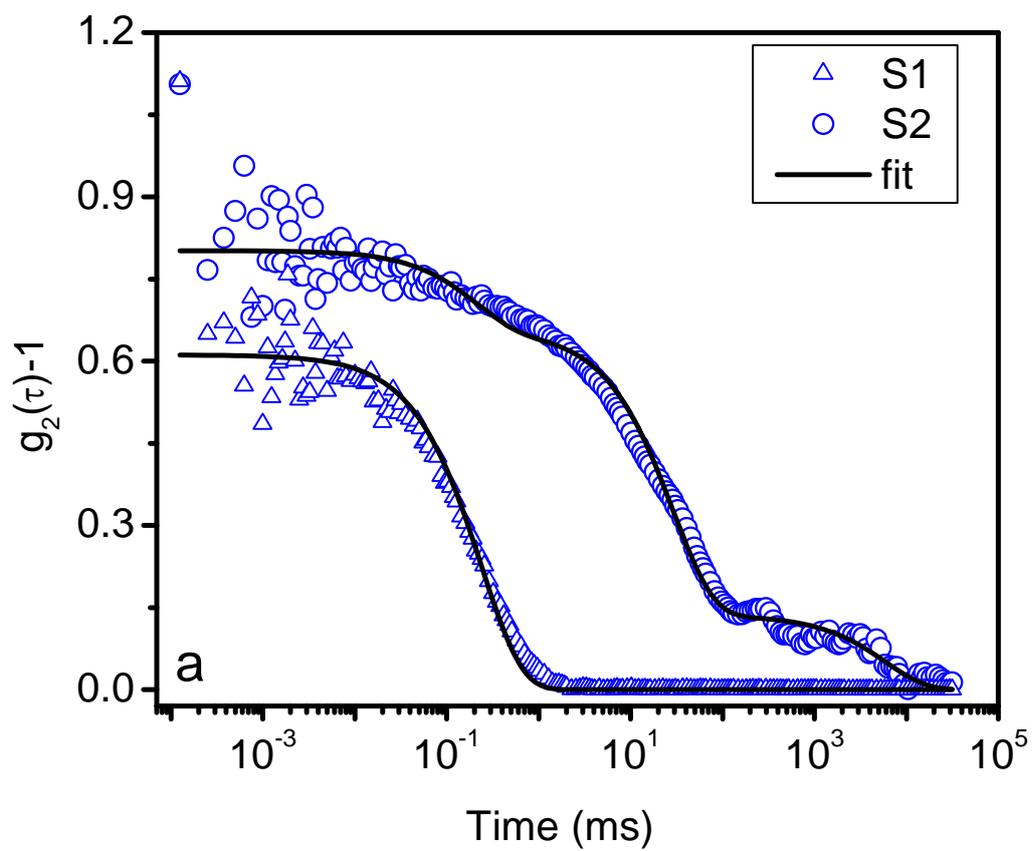

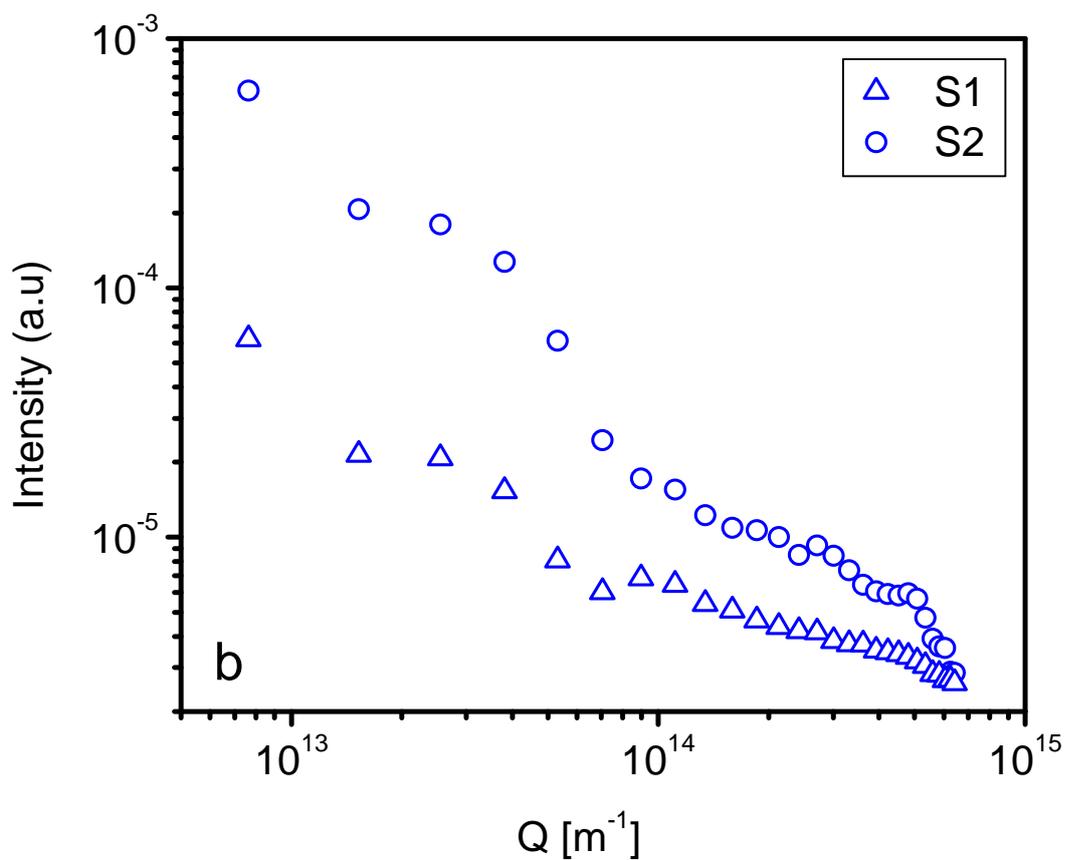

**Figure 1.** (a) The time correlation functions [$g_2(\tau)$-1] of aqueous polyacrylamide solutions obtained from the dynamic light scattering for samples S1 and S2. The data for S1 and S2 were fitted with a single and a sum of three exponential decay functions respectively. (b) Plot of the static light scattering intensity for two samples (S1 and S2) as a function of momentum transfer $Q$.

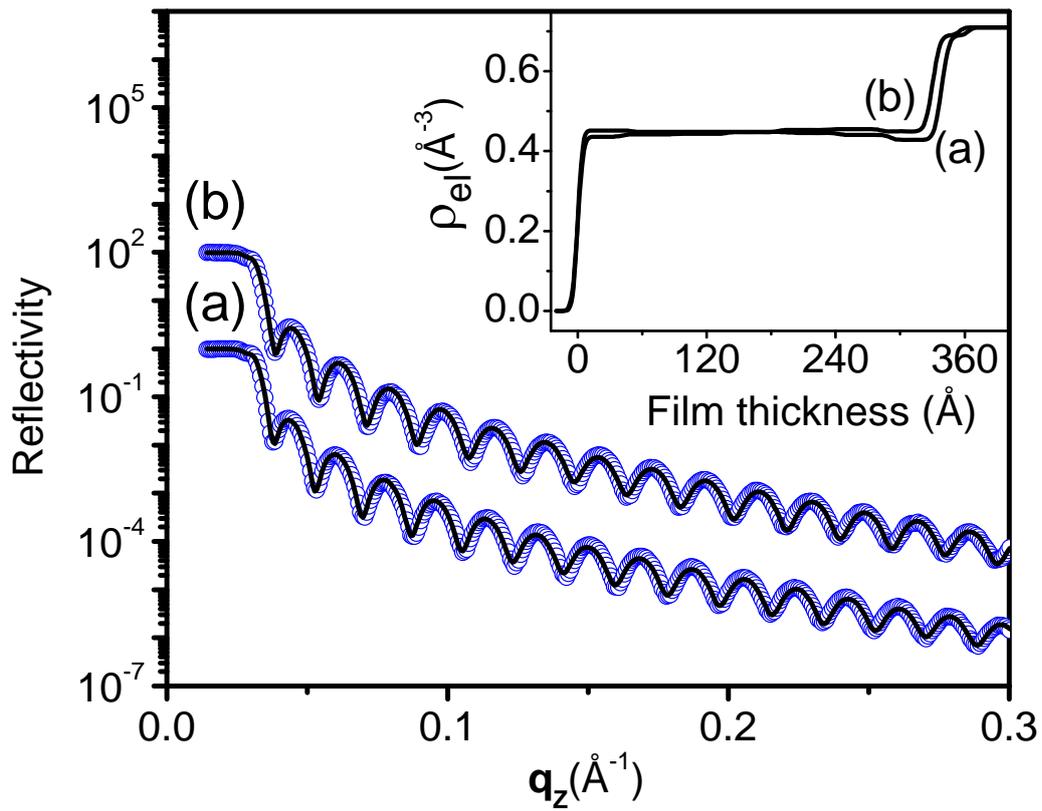

**Figure 2.** X-ray reflectivity data (symbols) with fitted profiles (lines) of as grown (a) and annealed film at 105°C (b) of a particular film of initial thickness 338Å. Inset shows corresponding electron density profiles.

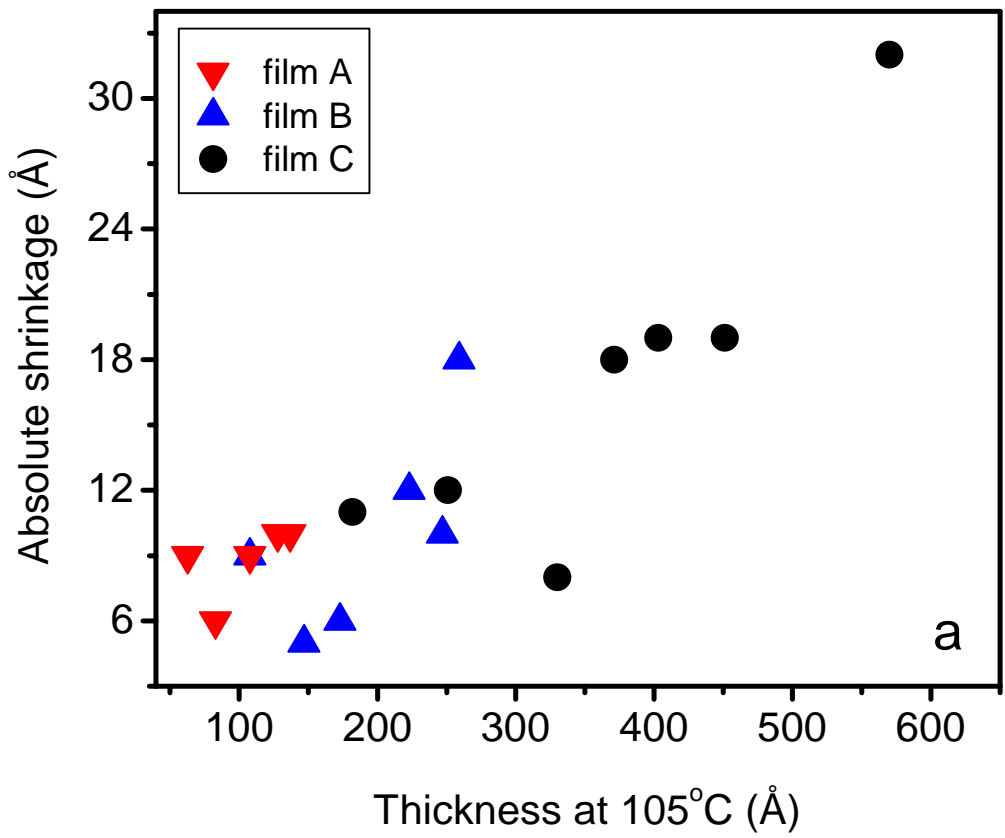

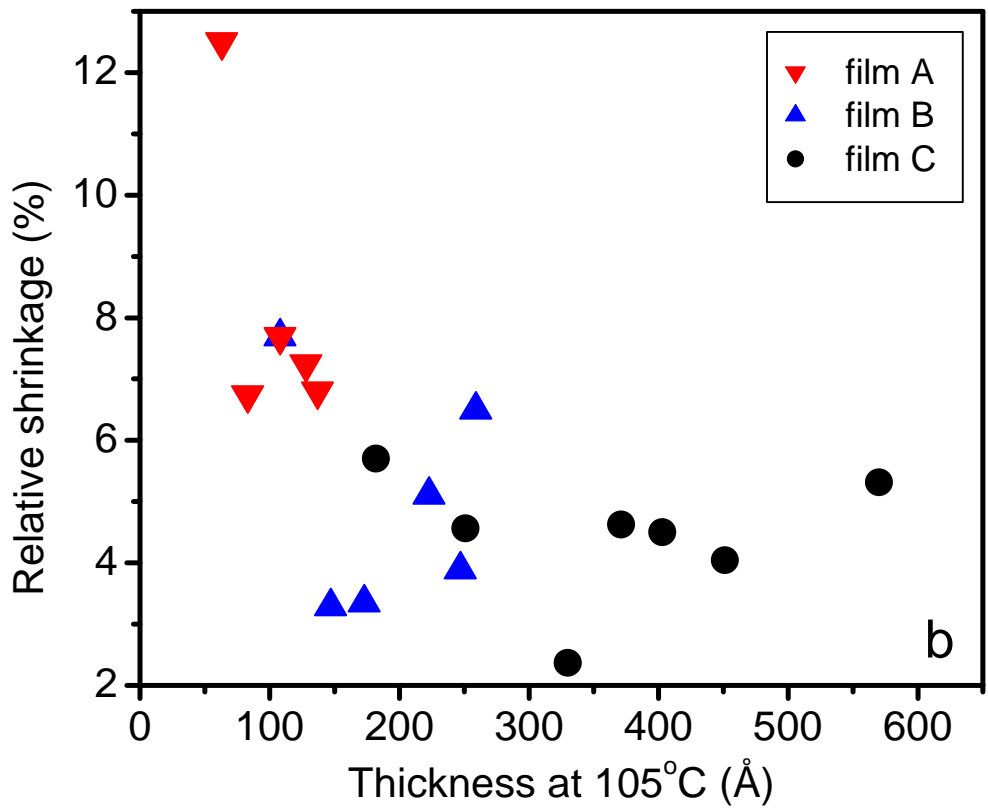

**Figure 3.** (a) Plot of absolute shrinkage for all the films as a function of film thickness at 105°C. (b) Relative shrinkage of the films as a function of film thickness at 105°C, calculated from the data in (a).

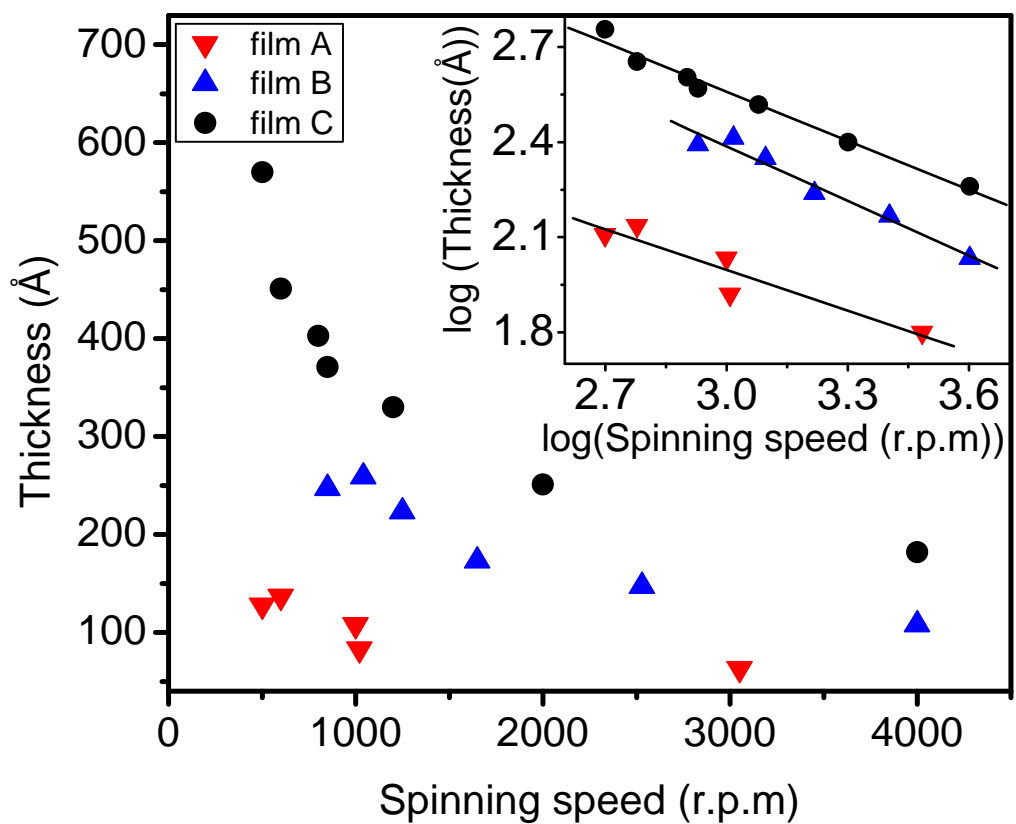

**Figure 4.** Film thickness as a function of spinning speed. Inset shows the logarithmic plots of the same data. The slopes of the straight line fits (*b*) in the inset represent the power law behaviors of the data.

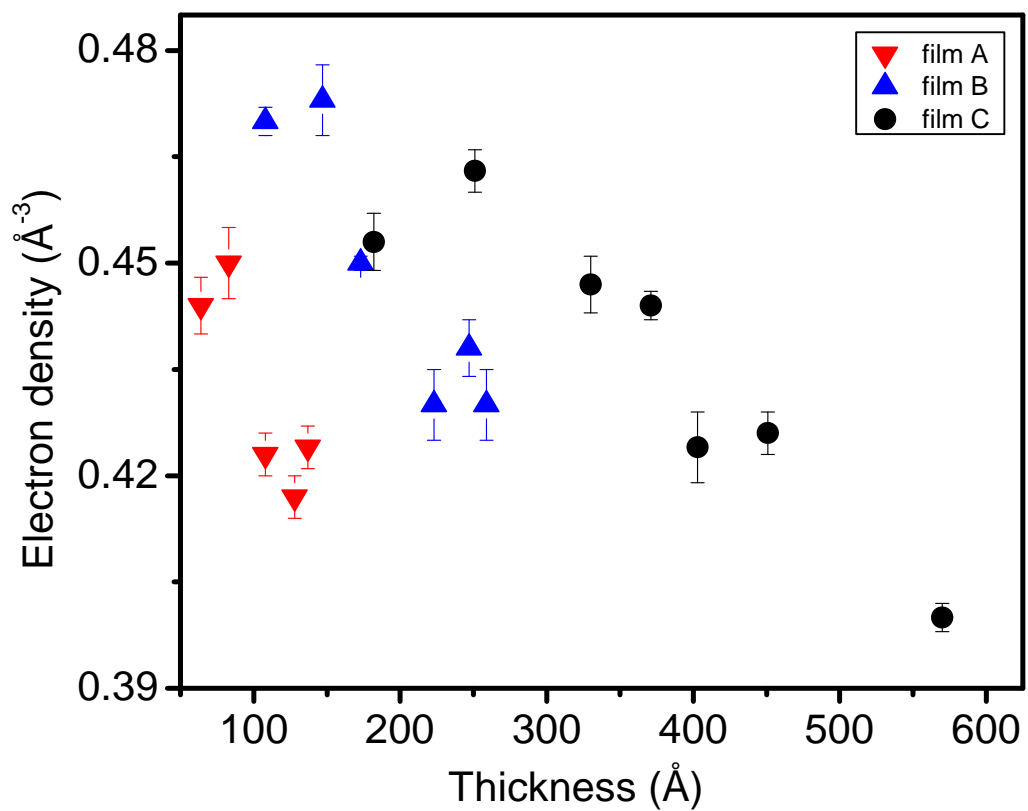

**Figure 5.** Average electron density of the films as a function of film thickness. The statistical errors are smaller than the size of the symbols. The error bars in the figure shows the variation of density when the least square fit value in the analysis of the reflectivity data is sacrificed by ± 1%.

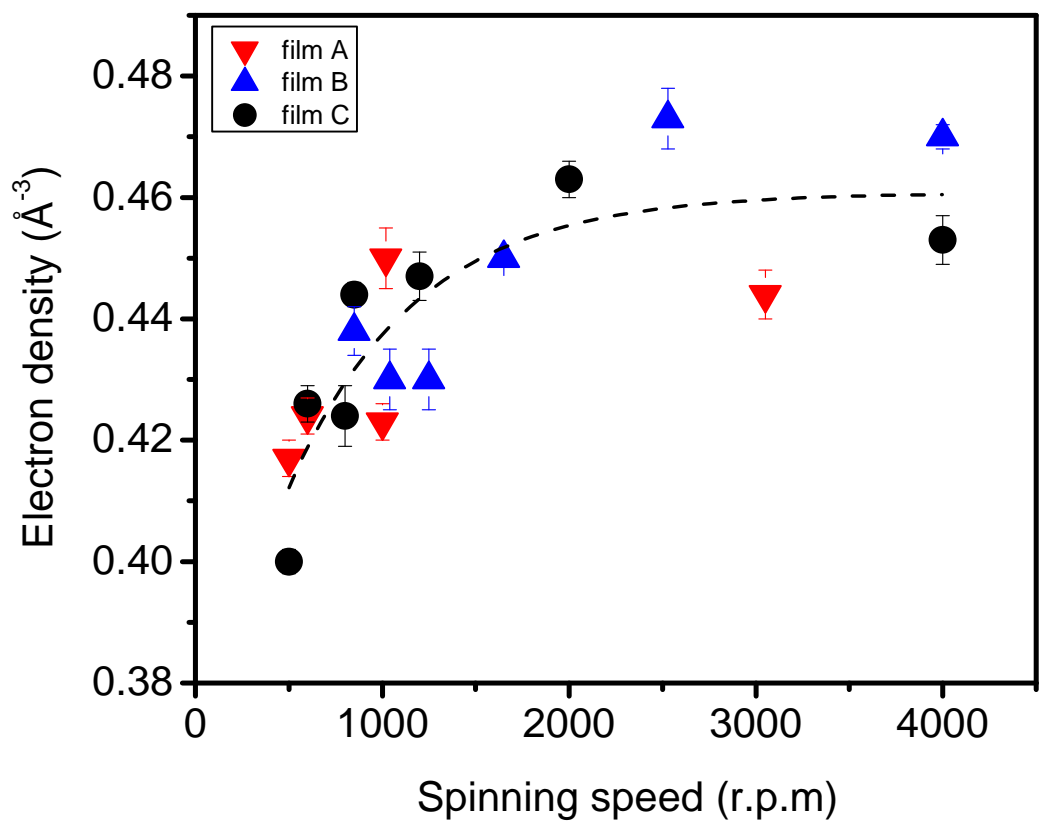

**Figure 6.** Average electron density of the films as a function of spinning speed. The error bars are same as described in the figure 5.

Table 1. Thicknesses and average electron densities of polyacrylamide films.

| film thickness at 105°C (Å) | film thickness at 165°C (Å) | average electron density at 105°C (Å$^{-3}$) | average electron density at 165°C (Å$^{-3}$) |
|---|---|---|---|
| 247 | 246 | 0.44 | 0.44 |
| 147 | 143 | 0.47 | 0.47 |
| 108 | 105 | 0.47 | 0.48 |

For Table of Contents use only

Title: Study of thickness dependent density in ultrathin water soluble polymer films

Authors: Mojammel H. Mondal and M. Mukherjee, K. Kawashima, K. Nishida, T. Kanaya

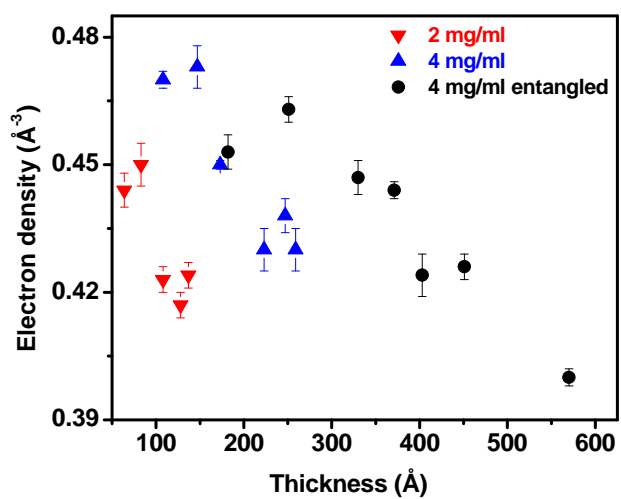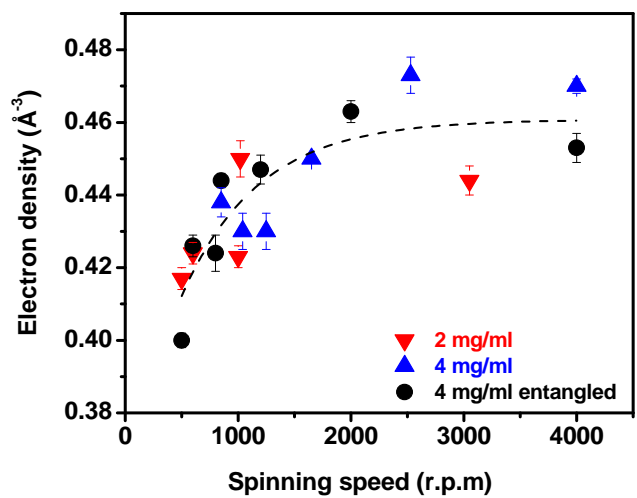

**Variation of electron density with thickness and spinning speed for polyacrylamide films**